\documentclass[conference]{IEEEtran}
\usepackage{amsthm}
\usepackage{amsfonts}
\usepackage{amssymb}
\usepackage{mathrsfs}
\usepackage{mathtools}
\usepackage{float}
\usepackage[pdftex]{graphicx}
\usepackage{amsmath}
\usepackage{color}
\usepackage{multirow}
\usepackage{graphicx}
\usepackage{graphicx,float,wrapfig,epstopdf,amsmath}
\usepackage{comment}
\usepackage{balance}
\usepackage{lipsum}
\usepackage[inline]{enumitem}
\usepackage{cuted}
\usepackage[caption=false,font=footnotesize]{subfig}
\usepackage{cite}
\usepackage{hyperref}
\usepackage{enumitem}
\newlist{myenumi}{description}{10}
\setlist[myenumi]{labelindent=\parindent, leftmargin=*, label=(\roman*), align=left}
\setlist[myenumi]{leftmargin=0pt}

\theoremstyle{definition}

\begin{document}


	\title{Federated Learning in Vehicular Networks 
	}

	\author{
		\IEEEauthorblockA{Ahmet M. Elbir$^{\dag}$, Burak Soner$^{\ddag}$,  Sinem \c{C}{\"o}leri$^{\ddag}$, Deniz G{\"u}nd{\"u}z$^{+}$ and Mehdi Bennis$^{++}$ }
		\IEEEauthorblockA{${\dag}$Department of Electrical and Electronics Engineering, Duzce University, Turkey\\
			${\dag}$Interdisciplinary Centre for Security, Reliability and Trust, University of Luxembourg, Luxembourg\\
			${\ddag}$Department of Electrical and Electronics Engineering, Ko{c} University, Istanbul, Turkey \\ 
			$+$Department of Electrical and Electronic Engineering, Imperial College London, U.K.\\
			$++$The Centre for Wireless Communications, the	University of Oulu, Finland}
		\IEEEauthorblockA{E-mail: ahmetmelbir@gmail.com, bsoner16@ku.edu.tr, scoleri@ku.edu.tr, \\
				 d.gunduz@imperial.ac.uk, bennis@ee.oulu.fi
			   }
	}
	
	
	%
	
	%
	%
	%
	
	\maketitle

	\begin{abstract}
		Machine learning (ML) has recently been adopted in vehicular networks for applications such as autonomous driving, road safety prediction and vehicular object detection, due to its model-free characteristic, allowing adaptive fast response. However, most of these ML applications employ centralized learning (CL), which brings significant overhead for  data transmission between the parameter server and vehicular edge devices. Federated learning (FL) framework has been recently introduced as an efficient tool with the goal of reducing  transmission overhead while  achieving  privacy through the transmission of  model updates instead of the whole dataset. In this paper, we investigate the usage of FL over CL in vehicular network applications to develop intelligent transportation systems. We provide a comprehensive analysis on the feasibility of FL for the ML based vehicular applications, as well as investigating  object detection  by utilizing image-based datasets as a case study. Then, we identify the major challenges from both learning perspective, i.e., data labeling and model training, and from the communications point of view, i.e., data rate, reliability, transmission overhead, privacy and resource management. Finally, we highlight related future research directions for FL in vehicular networks.

	\end{abstract}
	%

	
	\section{Introduction}
	As vehicles evolve with advanced safety features and  self-driving capabilities, massive amounts of data is generated by a variety of on-board sensors, such as camera, RADAR and LIDAR as well as proximity and temperature sensors. An autonomous vehicle is expected to generate about one gigabyte of data per second. However, currently, generated data is not systematically processed, stored, or analyzed for better inference. Recently, machine learning (ML) algorithms have been developed to learn from sensor measurements due to several advantages, including low computational complexity when solving optimization-based or combinatorial search problems and the ability to extrapolate new features from a limited set of features  contained in a training dataset.

\begin{figure*}[t]
		\centering
		{\includegraphics[draft=false,width=1.6\columnwidth]{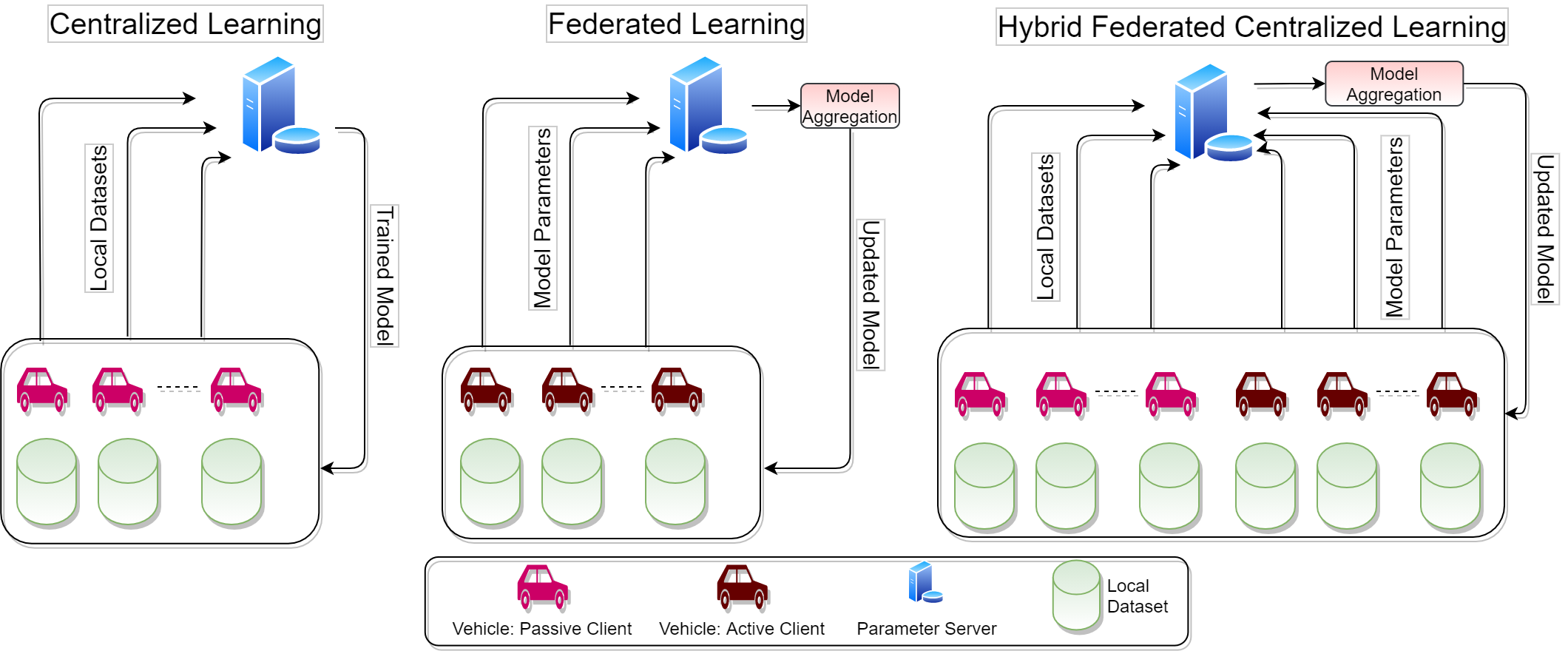} }
		\caption{Model training for CL, FL and HFCL in a vehicular network. 	
		}
		\label{fig_LearningModels}
	\end{figure*}

	The current trend in the usage of ML in vehicular networks focuses on centralized algorithms,  called centralized learning (CL), where a 	powerful learning algorithm, often a neural network (NN), is trained on a massive dataset collected from	the edge devices on the vehicles, as illustrated in Figure~\ref{fig_LearningModels}. The NN model  provides a non-linear mapping between the input, which contains  vehicle sensor data, and the output, which can be the labels of the sensor data. This mapping is learned by training the NN through the collection of the local sensor data from the edge devices, {\color{black}
		which is a supervised learning scheme.} Once model training is completed, the model parameters are sent back to the edge devices for prediction. {\color{black}
		However, the size of the generated data is huge when aiming at wider and deeper NN architectures for successful training. 
		Thus, training a model  with data transmission from the edge devices to the cloud center in a reliable manner may be too costly in terms of bandwidth, while incurring unacceptable delays, and infringing user privacy.}

	Federated learning (FL) has been recently introduced with the goal of bringing CL down to the edge~\cite{spm_federatedLearning}, as illustrated in Figure~\ref{fig_LearningModels}. In FL, instead of their local datasets, the edge devices only send the gradients of the learnable parameters derived from these local datasets to the cloud server. 	The cloud server aggregates these gradients and update the model parameters, which are then transmitted back to the edge devices. This procedure continues iteratively, until convergence. The training procedure is similar to  that of CL, except that FL does not involve the transmission of the whole dataset. This enables reducing both the complexity of ever growing datasets at the edge devices in the vehicles and the transmission overhead of these datasets to the cloud servers. 
	Hence, FL is a promising approach to efficiently train the learning models by preserving {the privacy of raw data} and reducing the transmission overhead in wireless communications.  While FL has already received great  interest within  wireless networks~\cite{spm_federatedLearning}, imparting FL to the vehicular networks is more challenging due to the dynamic nature of the channel characteristics of vehicular environments~\cite{Posner2021Feb,fl_veh_survey_OJCOM}. Furthermore,  previous works either  approach FL from only communication point-of-view~\cite{spm_federatedLearning,Posner2021Feb}, or suffer from investigating the use cases of FL in vehicular applications~\cite{fl_veh_survey_OJCOM}. The main contribution of this paper is to provide a comprehensive analysis of FL by considering both learning and communications aspects, based on the description and analysis of several learning-based vehicular applications. Moreover, we investigate the performance of FL in vehicular use cases, such as vehicular object detection and vehicle-to-infrastructure (V2I) beam selection based on image and LIDAR data.

	This paper aims to provide a comprehensive grasp on how vehicular networks  can benefit from  FL. First, we discuss  ML based vehicular network applications in the context of  vehicle management and traffic management. We present the advantages of  FL over CL for training the models in these vehicular network applications. By utilizing image- and LIDAR-based datasets, we present a case study of FL, such as 3D object detection. Then, we provide an extensive discussion on both FL- and communications-related research challenges  in a broad perspective. FL-related challenges include data diversity, labeling and model training, whereas communication-related challenges are transmission overhead, privacy, scheduling and resource allocation. Finally, we provide an extensive discussion on the major research issues and future research directions in making FL feasible in vehicular networks.

	\section{ML for Vehicular Applications}
	\label{sec:ML}
	ML-based techniques, particularly NN-based models, have become significantly effective in vehicular applications with the increase in the amount as well as diversity of   data generated by the sensors. Specifically, NN originates from the imitation of the human brain containing billions of neurons forming a neural network. There are mainly three types of ML, namely, supervised, unsupervised and reinforcement learning. In a supervised learning model, an NN is trained on a labeled dataset where an answer key is provided beforehand. In contrast, unsupervised learning studies the clustering of the unlabeled data by exploiting the hidden features/patterns derived from the dataset.  Reinforcement learning (RL) also uses unlabeled data and the model parameters are learned based on the award and penalty mechanism, which are formulated as a function of varying environment characteristics. 

	In the following, we present several  ML-based vehicular applications and the associated challenges in two categories: vehicle management and traffic management.
	
	\subsection{Vehicle Management: Autonomous Driving}
\begin{figure}[t]
		\centering
		{\includegraphics[draft=false,width=\columnwidth]{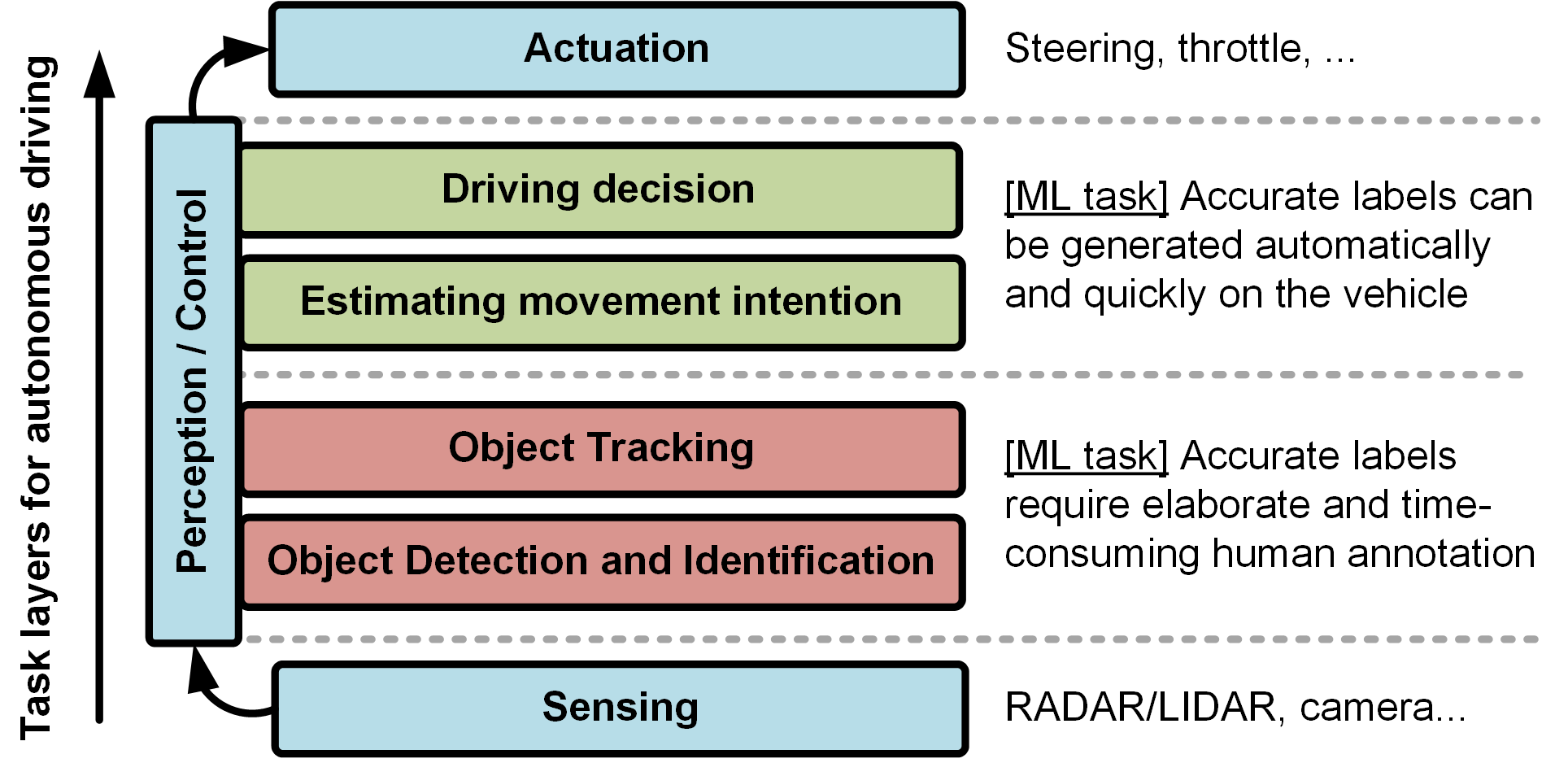} }
		\caption{The hierarchical structure of tasks for autonomous driving. 
		}
		\label{fig_autoDrvLayers}
	\end{figure}

	The aim of autonomous driving is to navigate a vehicle through a road environment without collisions. To achieve this, the vehicle needs to detect, identify, localize and track surrounding objects, such as pedestrians, trees and other vehicles relative to its own frame of reference, and adjust its driving dynamics accordingly. {\color{black} The layered nature of this end-to-end autonomous driving task, depicted in Figure~\ref{fig_autoDrvLayers}, renders conventional methods that use an ensemble of hand-crafted computer vision techniques for each layer sub-optimal. Deep neural networks that are trained in an end-to-end manner on a huge amount of RADAR/LIDAR/camera sensor data, such as convolutional neural networks (CNNs), have emerged in recent years as a better alternative~\cite{fl_veh_survey_OJCOM}.} 
	%

	In~\cite{ML_autonomous}, a human-like decision-making method is proposed for autonomous driving by using CNNs. The input of CNN is the LIDAR data collected from multiple vehicles to provide depth information, whereas the {\color{black} output of the }CNN is the decision regarding the speed and steering of the vehicle. {\color{black} While this demonstrates the advantage of using 
		trained CNNs for autonomous driving, CNNs require a large amount of precisely labeled data for accurate prediction.} Therefore, the training is usually conducted in a cloud  server in an offline manner.	The main drawback in offline training is that the trained model cannot adapt to the  environment dynamics. 
	The usage of FL can provide the adaptation to the environmental changes, such as feature learning in different geographical locations. However,  FL utilizes gradient computation at the edge device, which necessitates immediate and accurate labeling of online data. Therefore, the higher layers of the autonomous driving ML application stack shown in Figure~\ref{fig_autoDrvLayers}, such as intent estimation and driving decisions 
	can get higher benefit from FL strategies rather than the lower layers such as object detection and tracking. 
	

	\subsection{Traffic Management: Infotainment and Route Planning}
	{\color{black}Infotainment and route planning are crucial for preventing traffic jams, sustaining an efficient distribution of resources during normal traffic, and enabling effective emergency response during extreme situations.} Data-driven ML solutions are superior to conventional model-based approaches for these applications since they can easily adapt to the random changes in the system dynamics over time caused by human involvement, which can be difficult to model mathematically. {\color{black}For instance, in \cite{sdn_infotainment}, a deep learning based resource management scheme, which learns the hidden patterns in data traffic over a vehicular software-defined network (SDN), is shown to improve packet traffic efficiency compared to the conventional methods. Similar solutions exist for route planning and efficient management of vehicle traffic.} While these applications benefit from versatile ML solutions, they rely on frequent model retraining due to the ever-changing dynamics of the systems.
	

	\section{FL for Model Training in Vehicular Networks}
	\label{sec:FL}
	
	{\color{black}Conventional approaches for training ML models in vehicular applications consider a central server, which collects the raw data from vehicular edge devices, computes gradients according to the current state of the model and the incoming data, then updates the model parameters accordingly. However, this approach cannot easily yield a model that can adapt to local changes as required by the applications. Many of the nodes in a given vehicular network cannot comply with the requirements for high transmission overhead and raw data privacy due to the highly dynamic and harsh communication channel, hindering the local adaptation capability of the model, since the local data from those nodes cannot participate in the model training to better represent the distribution of the global dataset. Decentralized training with  FL scheme enables adaptation to the data distribution since the gradients are computed at the edge devices based on  local datasets, and the model updates themselves, instead of the much larger raw data, are transmitted to the central server. This not only reduces the transmission overhead, but also provides some level of privacy. Thus, it enables more nodes to participate in model training, which results in a model that better adapts to the local changes inside the global dataset.
		
		The proposition of FL, which enables it to adapt to local changes, is based on the ``mini-batch learning'' technique in conventional ML model training, where the dataset is partitioned into smaller sub-blocks that are used for parameter updates rather than the whole dataset. The gradients are computed for each of these sub-blocks, i.e., mini-batches, and then a combination of these gradients, typically the average, is used as the gradient value for a parameter update step, which is repeated iteratively until convergence. Since the computation of the gradients for different mini-batches are independent, FL schemes can exploit the local processing capabilities of the edge devices in the vehicles to compute and transmit these in parallel, as illustrated in Figure 2. The central server then simply combines the collected gradients and performs model  update, then transmits the new model parameters to each vehicle. This reduces the transmission overhead and preserves privacy to some extend since the much larger and vulnerable raw data is not transmitted, enabling significantly more nodes to participate in the model training, thus, increasing the local adaptivity of the ML model.
		
		Although FL has its advantages over CL, not all applications can benefit from FL strategies for model training. Using FL is advantageous for applications and environments that satisfy three conditions:

		\begin{figure*}[h]
			\centering
			\subfloat[]{\includegraphics[draft=false,width=.7\columnwidth]{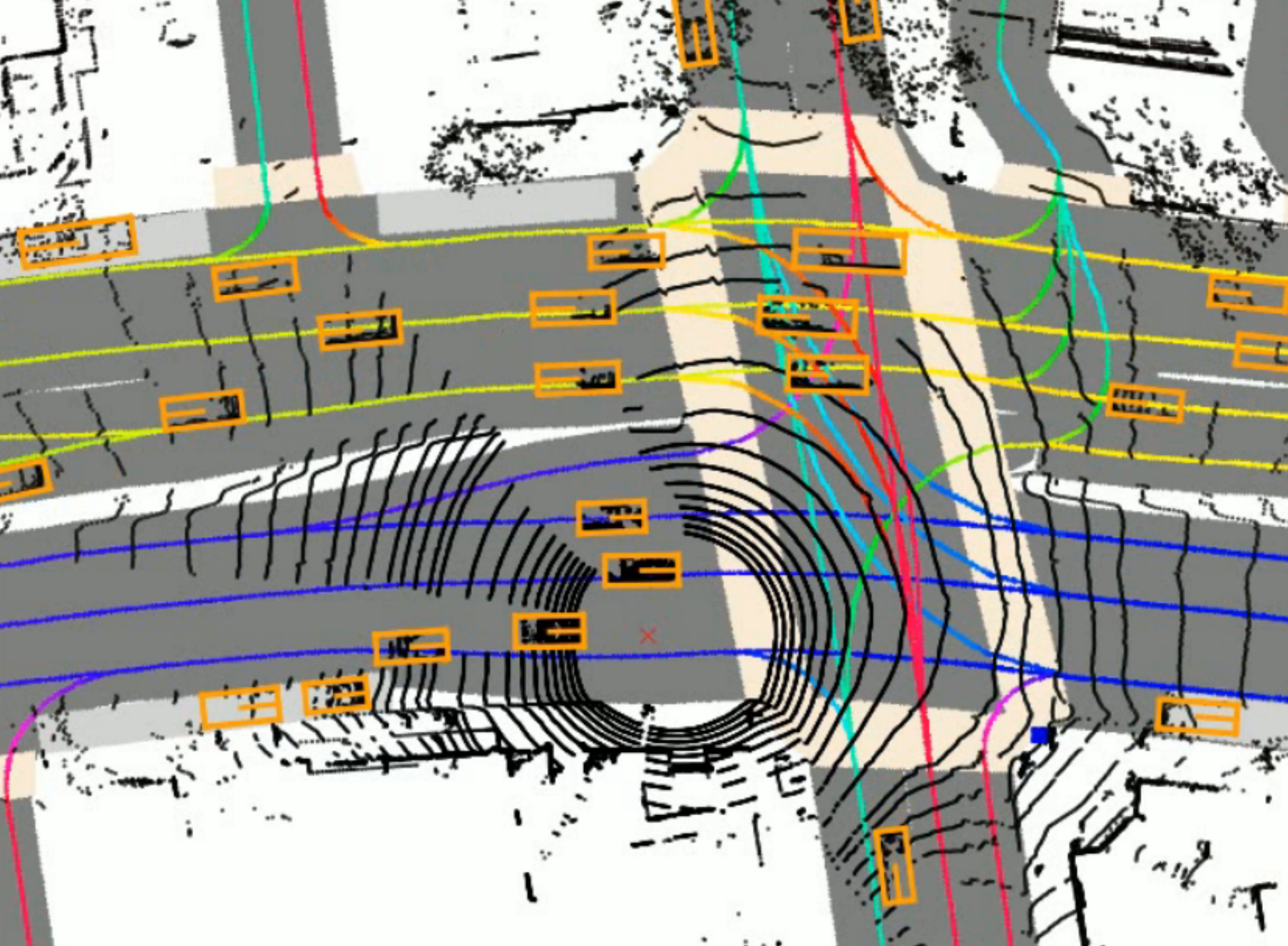} \label{fig_AVa}}
			\subfloat[]{\includegraphics[draft=false,width=.75\columnwidth]{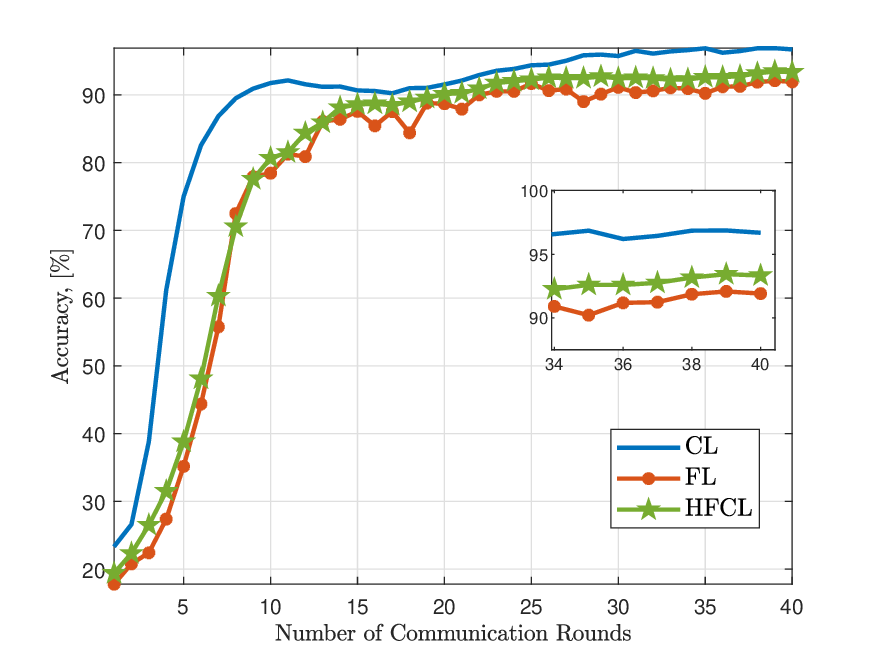} \label{fig_AVb}}
			\caption{ (a) Visualization of input and output data, and (b) learning accuracy for 3D object detection.
			}
			\label{fig_AV}
		\end{figure*}

		\begin{enumerate}
			\item The application requires a model that adapts to new conditions, which constantly change on the edge, and retraining/update-from-server/deployment for each edge device needs to happen fast.
			\item Data gathered by the edge device cannot be off-loaded to a cloud server within the required time frame, either because of its sheer volume, or due to privacy concerns
			\item Data gathered by  edge devices can be quickly and accurately labeled on the same device for gradient computations.
		\end{enumerate}
		As a new emerging field still in its infancy, there are a very limited number of works considering FL in vehicular applications that satisfy these three conditions~\cite{FL_vanet_Bennis,FL_vnet1_edgeComp}. {\color{black}In \cite{FL_vanet_Bennis}, FL has been considered in a vehicular network, where the communication between the data center and the edge devices is assisted by road side units to ensure low latency for gradient data transmission. Specifically, a Lyapunov optimization based approach is proposed to minimize the delays incurred by the transmission of gradient data in FL. In \cite{FL_vnet1_edgeComp}, the authors propose a selective model aggregation approach, where the data center collects the gradient data from only ``fine'' edge clients in a vehicular network, such as the devices with high data quality and transmission power capability. Then, the performance of FL is compared to that of CL based on both MNIST and BelgiumTSC image classification datasets, which are composed of the images of the numbers and traffic signs, respectively.} While these works demonstrate the above-mentioned advantages of using FL over CL for a constrained set of applications, they are not directly applicable for realistic ML-based vehicular applications. FL still faces significant open challenges for use in such realistic applications like autonomous driving and traffic management via infotainment and route planning.}

	As discussed above, FL is advantageous as compared to CL in terms of communication overhead while FL requires computational resources from the edge devices to perform model computation.  Therefore, not all the vehicles can participate in training if they do not have enough computational resources. To address this issue, \cite{elbir2021HFCL} devised a hybrid federated and centralized learning (HFCL) framework, in which only the vehicles with sufficient resources employ FL, while the remaining ones send their local datasets to the PS, which computes the model parameters on behalf of them (see Figure~\ref{fig_LearningModels}). Then, the model parameters corresponding to all vehicles are aggregated at the PS. Thus, a part of the vehicles perform FL while the remaining ones employ CL, which demands dataset transmission prior to the training. As a result, HFCL provides a trade-off between the higher learning accuracy of CL and the lower communication overhead of FL at the expense of sacrificing privacy. The privacy concerns in HFCL can be addressed through coded FL~\cite{Prakash2020Nov}.

	%
	%

	\subsection{Case Study 1: 3D Object Detection}
	As a case study, we evaluate the performance of learning frameworks, e.g., CL, FL and HFCL, on 3D object detection problem in vehicular networks, based on the Lyft Level 5 AV dataset~\cite{elbir2021HFCL}, collected from LIDAR equipment and cameras mounted on vehicles. The input data is selected as a top view image of the ego vehicle, which includes the received LIDAR signal strengths for different elevations, and the output is the classified representation of the vehicles/objects as boxes, which is obtained by the preprocessing of the images from the cameras, as illustrated in Figure~\ref{fig_AVa}. The training dataset is collected from $10$ vehicles in different areas after preprocessing of camera and LIDAR data. We assume $3$ out of $10$ the vehicles are passive while the remaining ones are active clients. Each local dataset includes $10^3$ input-output pairs, whose sizes are $336\times 336\times3$ and $336\times 336\times1$, respectively. Hence, the total number of data symbols is  $(336\times 336\times3 + 336\times 336)\times 10^4 \approx 4.5\times 10^9$. The dataset has $9$ classes, i.e., $\mathrm{car}$, $\mathrm{motorcycle}$, $\mathrm{bus}$, $\mathrm{bicycle}$, $\mathrm{truck}$, $\mathrm{pedestrian}$, $\mathrm{other}\hspace{2pt} \mathrm{vehicle}$, $\mathrm{animal}$, $\mathrm{emergency}\hspace{2pt} \mathrm{vehicle}$, which are represented by the boxes as shown in  Figure~\ref{fig_AVa}.  For 3D object detection and segmentation, we have used \texttt{U-net}~\cite{unet} with $8$ convolutional layers to learn the features of the input data. further,  the total number of parameters in \texttt{U-net} is approximately $2\times 10^6$, and  $40$ communication rounds for model training are carried out~\cite{elbir2021HFCL}.
	
	The training performance of the all methods is presented in Figure~\ref{fig_AVb}, from which we observe that the HFCL provides a moderate performance between CL and FL. Nevertheless, all of the vehicles can participate in training while  conventional FL methods cannot support it whereas the transmission overhead of CL is prohibitive. Thus, HFCL solves the trade-off between the computational capability of the clients and the transmission overhead. The overhead of CL is due to the transmission of whole data symbols, i.e., approximately $ 4.5\times 10^9$. In contrast, the complexity of FL is due to the two way (edge $\leftrightarrows$ server) transmission of the model updates  during training until convergence, i.e., $2\times40\times  (2\times 10^6) = 160\times 10^6$ for $40$ iterations. As a result, FL and HFCL have approximately $28$ and $3$ times lower transmission overhead as compared to CL, respectively. 
	

	%
	%
	%
	\subsection{Case Study 2: Millimeter-wave Beam Selection}
\begin{figure}[t]
		\centering
		{\includegraphics[draft=false,width=.7\columnwidth]{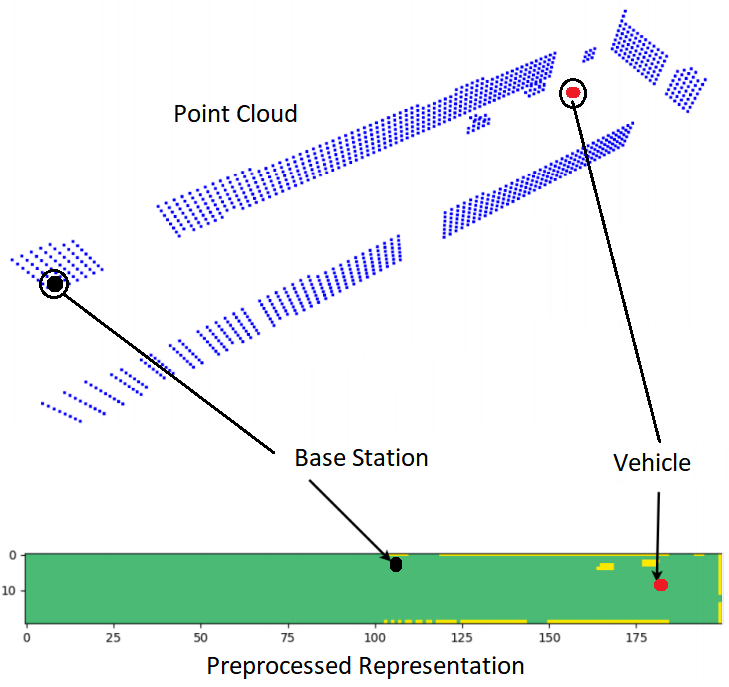} \label{fig_LIDARa}}
		\caption{ Visualization of the LIDAR point cloud data for a single vehicle moving on a road
			for mm-Wave beam training~\cite{fl_deniz_beamTraining}.
		}
		\label{fig_LIDAR}
	\end{figure}

	In this scenario, the performance of FL is evaluated for vehicular to infrastructure (V2I) millimeter-wave (mm-Wave) communication~\cite{fl_deniz_beamTraining}. In particular, for a vehicle moving on a road, a laser scan of the environment is obtained via LIDAR equipment to obtain the beam directions with the best mm-Wave channel conditions to establish communication with a base station (BS) on the road. The LIDAR dataset is collected for $10$ vehicles in a similar scenario, for which the input of the learning model is the LIDAR point cloud measurement, as shown in Figure~\ref{fig_LIDAR}, and the output is the index of best beam directions. In order to construct the non-linear  mapping between the LIDAR  data and the beam directions, a CNN with $8$ layers and $7462$ learnable parameters is trained for $20$ epochs, then the trained models for FL and CL are deployed for beam selection. The results show that FL and CL have $88.6\%$ and $89.1\%$ beam selection accuracy to select the best available mm-Wave beams between the vehicle and the BS~\cite{fl_deniz_beamTraining}. While a slight performance loss is observed in FL, it has approximately $300$ times lower transmission overhead as compared to CL. The effectiveness of FL is due to the transmission of only the model parameters, while CL involves the transmission of the whole dataset.

	\section{Research Challenges and Future Directions}

	\subsection{FL-Related Research Challenges}
	
	
	\subsubsection{Data Diversity}
	In FL,  training data is located at the edge devices, which causes data diversity due to the non-uniform distribution of the datasets at the edge devices.	For example, in autonomous driving scenario, the image data obtained from vehicles in different locations have different distributions. Data diversity  causes large variances in the averaged gradient data, and therefore, decreases the convergence rate for the learning models. {\color{black}
		For example, the features of the collected image data in different locations {\color{black}increases} the diversity of the dataset, which {\color{black}makes} NN unable to perform  feature-extraction and feature-representation well. Furthermore, the number of edge devices in vehicular networks is smaller than general wireless networks due to the large distance among the vehicles. This leads to the model aggregation from fewer number of edge devices, which also makes model training a challenging task in vehicular applications.}  To improve the model training performance against data diversity caused by either non-uniform data distribution or insufficient number of edge devices, one possible solution is to increase the model size, i.e., enlarging the width and the depth of the NN model so that the NN can provide robust feature representation~\cite{elbir2020FL}.
	
	%
	%
	%
	
	\subsubsection{Labeling}{\color{black}
		ML techniques are mostly supervised, i.e., the dataset is labeled. For example, an image dataset of vehicles includes labels, such as ``car'', ``motorcycle'', ``truck'' (see Figure~\ref{fig_AV}). Data labeling demands certain amount of labor to label/annotate the collected training data, which is usually done in an  offline manner in current commercially available autonomous vehicle prototypes. In FL based training, the dataset should be labeled so that each edge client can compute the model updates based on the local labeled dataset. This is one of the main challenges of FL in major learning based vehicular applications, such as autonomous driving and object detection.  A possible approach to resolve this problem is to use reinforcement learning (RL) techniques~\cite{rl_Autonomous_withoutLabel}, which do not require labeled data. In particular, RL is based on the award and penalty mechanism, which are formulated as a function of varying environment characteristics to optimize the objective of the learning problem. The main challenge in RL is that RL requires longer training times and the performance is usually worse than the supervised techniques due to the absence of labels. }
	
	\subsubsection{Efficient Model Training}
	The efficiency of model training can be improved by the use of transfer learning  (TL) based approaches. TL is an ML method, where a model developed for a certain task is reused as the starting point for a model on a different task. In~\cite{elbir2020TL}, TL has been proposed for cognitive RADAR applications, where an ML model is used for different sensor selection tasks. The application of TL for vehicular networks is advantageous. For instance, instead of training a model from scratch, a well-trained model with large dataset can be used with a soft parameter update for smaller datasets. In this case, the parameter update involves lower complexity since only a small portion of the NN is trained, which leads to more efficient  model training.	The success of TL in vehicular networks depends on data similarity. Specifically, the TL accuracy strongly relies  on the similarity between the newly collected data at the edge devices in the vehicular network and the training data used to   pre-train the  model. In order to obtain higher accuracy from the NN with new data, a larger portion of the model should be updated. Thus, there is a trade-off between the similarity/diversity of the datasets and the required training complexity. While data similarity/diversity issue has been studied for cognitive RADAR applications~\cite{elbir2020TL}, its effect on vehicular applications has not been exploited. In addition, the diversity of the datasets can incur difficulties when performing TL due to non-uniform distribution of the dataset, which has been studied in~\cite{FL_Gunduz}, accommodating a shallow ML model without the focus on TL. As a result, new approaches need to be developed to make FL model training more feasible in vehicular network applications.
	
	
	\subsection{Communications-Related Research Challenges}

	\subsubsection{{\color{black}Transmission Overhead}}
	Compared to the CL based techniques, FL allows us to reduce the transmission overhead by replacing raw data transmission with model update parameter transmission. However, the size of the model update parameter set is directly proportional to the size of the learning model. Thus, the transmission of model parameters may become a bottleneck if the model involves massive number of learnable parameters.
	
	
	{\color{black} The solution is either to reduce the transmission overhead or increase the capacity of the channel.} There are various methods  to reduce the transmission overhead in FL-based framework including sparsification and quantization. The sparsity of the gradients, i.e., most of the gradients being zero, can be exploited to reduce the amount of transmitted data~\cite{FL_Gunduz}. While this approach reduces the transmission overhead, it increases encoding/decoding complexity at the parameter server and edge devices in a vehicular network for reliable performance. 
	
	{\color{black}Alternatively, channel reliability can be improved to facilitate faster convergence to more accurate models. For example, the authors in \cite{FL_Gunduz} attempt to  design an FL-friendly communication protocol from scratch based on the exploitation of  the broadcast/multiple-access nature of the downlink/uplink channels in an FL framework.} Using analog links in noisy band-limited channels for both uplink and downlink rather than digital links has been demonstrated to provide higher reliability at lower transmit power, especially when the local data among edge devices is not uniformly distributed.  {\color{black}Development of environment-aware heterogeneous architectures combining the strengths of different standard-compliant vehicular communication technologies, e.g., choosing between IEEE 802.11p, millimeter-wave and visible light communications based on road/channel conditions, is a promising  research direction for improving the overall network reliability of FL-based vehicular frameworks.} 
	
	%
	
	\subsubsection{{\color{black}Security and Privacy}}
	During FL training, different types of devices may participate in the learning stage. Thus, untrusted devices can join the network more easily, which brings security and privacy issues. The security and privacy of the devices in the network can be achieved through the use of reputation management (reward and punishment) based approaches. In~\cite{FL_vnet1_edgeComp}, authors propose a method, where each edge device receives a reward in exchange for their computation of power and data quality. However, \cite{FL_vnet1_edgeComp} considers a simple FL framework with a single server. In a realistic vehicular network scenario, there can be multiple access points acting like servers in FL, increasing the dimension of the reputation management problem. As a result, their usage in vehicular networks requires further research for multi-server FL architectures.
	
	\subsubsection{{\color{black}Scheduling and Resource Management}}
	{\color{black}The availability of wireless communication resources and the packet error performance of each node in a wireless network vary greatly among nodes due to both device heterogeneity and spatial distribution.} {\color{black}Since FL convergence rate is directly affected by the performance of the communication link between each edge device and the central FL controller, scheduling and resource management of nodes participating in an FL scheme need to be explicitly optimized with the objective of maximizing the FL convergence rate. However, this problem becomes ill-defined for vehicular networks where the links between the vehicular edge devices on the road are usually intact whereas the links to the central controller from each device experiences frequent and sporadic handovers and drop-outs. A ``collaborative'' FL solution where the vehicular edge devices instead utilize multiple hops between neighboring vehicular edge devices to reach the central controller can decrease the sensitivity of the FL framework performance to such network-related shortcomings. While such a framework is proposed in \cite{collabFL}, it is specifically tuned to the network topology and dynamics of regular wireless/cellular networks, which are significantly different from vehicular networks. Therefore, further research on scheduling and resource management strategies for FL, focusing on the low-density / high-mobility network topology in the vehicular environment, is needed.}
	
	
	\section{Summary}
	In this paper, we presented the FL-based framework {\color{black}for decentralized training of ML models}, as an efficient learning scheme for vehicular networks and edge intelligence, in {\color{black} contrast to  classical ML techniques based on CL}. We enlisted several applications of vehicular networks that are particularly amenable to FL, such as autonomous driving, infotainment and route planning. We identified a number of  major research challenges for FL in vehicular networks, involving both learning challenges, i.e., data labeling and model training, and  the communications challenges, i.e., data rate, reliability, transmission overhead, privacy and resource management, along with the related future research directions.

		\section*{Acknowledgment}
	This work was supported by CHIST-ERA grant CHIST-ERA-18-SDCDN-001, and the Scientific and Technological Council of Turkey 119E350.

	\bibliographystyle{IEEEtran}
	\bibliography{IEEEabrv,references_overview1}
	\balance

\end{document}